\documentclass[aps,prl,twocolumn,superscriptaddress,longbibliography]{revtex4-1}

\usepackage[colorlinks=true, allcolors=blue]{hyperref}
\usepackage{upgreek} 
\usepackage{natbib}
\usepackage{graphicx}

\begin{document}

\title{Nuclear charge radii of Na isotopes: A tale of two theories}

\author{B. Ohayon}\email{Corresponding author: bohayon@ethz.ch}
\affiliation{
Institute for Particle Physics and Astrophysics, ETH Z\"urich, CH-8093 Z\"urich, Switzerland 
}

\author{R.F. Garcia Ruiz}
 \affiliation{Massachusetts Institute of Technology, Cambridge, Massachusetts 02139, USA
}

\author{Z.~H.~Sun} 
\affiliation{Physics Division, Oak Ridge National Laboratory,
Oak Ridge, Tennessee 37831, USA}
\affiliation{Department of Physics and Astronomy, University of Tennessee,
Knoxville, Tennessee 37996, USA} 

\author{G.~Hagen} 
\affiliation{Physics Division, Oak Ridge National Laboratory,
Oak Ridge, Tennessee 37831, USA}
\affiliation{Department of Physics and Astronomy, University of Tennessee,
Knoxville, Tennessee 37996, USA} 

\author{T.~Papenbrock}
\affiliation{Department of Physics and Astronomy, University of Tennessee,
Knoxville, Tennessee 37996, USA} 
\affiliation{Physics Division, Oak Ridge National Laboratory,
Oak Ridge, Tennessee 37831, USA} 

\author{B. K. Sahoo}\email{bijaya@prl.res.in}
 \affiliation{
Atomic, Molecular and Optical Physics Division, Physical Research Laboratory, Navrangpura, Ahmedabad 380058, Gujarat, India
}

\date{\today}

\begin{abstract}
The accuracy of atomic theory calculations limits the extraction of nuclear charge radii from isotope shift measurements of odd-proton nuclei. For Na isotopes, though precise spectroscopic measurements have existed since more than half a century, calculations by different methods offer a wide range of values. Here, we present accurate atomic calculations to reliably extract the Na charge radii.
By combining experimental matter radii with nuclear coupled-cluster calculations based on nucleon-nucleon and three-nucleon forces, we constrain the parameters obtained from the atomic calculations.
Therefore, this study guides atomic theory and highlights the importance of using accurate atomic and nuclear computations in our understanding of the size of light nuclei.

\end{abstract}

\maketitle


\textit{Introduction.---}
The understanding of the evolution of the size of a nucleus with extreme numbers of protons and neutrons is a challenge for microscopic nuclear theory \cite{Cau01,Lap16,Lon18,eks15,soma2020,sam2020,Soma2021}. In recent years, simultaneous developments in experimental techniques, as well as the atomic and nuclear theory, have provided great advances in our understanding of the nuclear size away from stability \cite{mil19,gar16,gro20,Koszorus2021}. Accurate nuclear charge radii calculations with quantifiable uncertainties are now becoming available for the light and medium mass nuclei, enabling detailed comparisons with experimental data \cite{Lap16,Lon18,sam2020,2021-Al, Koszorus2021}. %
Light nuclear systems with mass number $A<40$, whose properties can now be assessed by different \textit{ab initio} many-body methods \cite{Lu13,Lon18,Lap16,2019-Boron,sam2020,soma2020}, are a testing ground for nuclear theory.
However, nuclear charge radii measurements of these systems are scarce, with most of the available data obtained from spectroscopic measurements of atomic isotope shifts (ISs) \cite{2012-Yor,2021-Al, Koszorus2021}, whose uncertainties are dominated by calculated atomic parameters.
The reason for this is that in light elements, the nuclear volume contributions to the ISs, which allows a measure of the charge radius, decrease exponentially with atomic number, leaving the mass shifts (MSs), strongly affected by electronic correlations \cite{2021-Al}, to dominate. 
Hence, the accuracy of atomic theory is one of the current main limitations for extending our knowledge of the nuclear size in this frontier region of the nuclear chart. 

IS measurements cannot be compared directly with atomic calculations. Where at least three stable isotopes exist, independent charge radii measurements determined by muonic x-ray transition energies or elastic electron scattering \cite{1995-Fricke} can be used in combination with IS measurements to benchmark atomic calculations. However, this procedure cannot be applied to most odd-proton elements as they have only one stable isotope.
The isotopes of Na ($Z=11$) provide a distinct example of how joint developments in both atomic and nuclear theory are critical to guide our understanding of the evolution of the nuclear size. The IS measurements for these isotopes have existed since more than four decades~\cite{1978-Na2}. However, it has been a major challenge to perform accurate atomic calculations required to extract the nuclear charge radii values from the experimental data.
For this reason, Na isotopes are some of the rare cases where matter radii are known with higher precision than charge radii \cite{Su98,2014-NaRM}.

In this Letter, we report on new accurate relativistic atomic calculations, with quantifiable uncertainties, that enable the extraction of charge radii values for the Na isotopic chain. The available experimental data on matter radii alongside the recent progress of nuclear theory are used to establish constraints on atomic parameters, providing guidance to the developments of atomic many-body theory.

\textit{Isotope shifts.---} Changes in the root-mean-squared nuclear charge radii, $\delta r_c^2$, can be inferred from measurements of ISs, $\delta \nu$, using the linear expression \cite{1984-King}
\begin{equation}\label{eq:KP}
    \delta \nu ^{A',A} =  K (1/M_{A'}-1/M_A) +  F (\delta  r_c^2)^{A',A},
\end{equation}
where $K$ and $F$ are the transition-dependent MS and field-shift (FS) constants, respectively, which are to a good approximation isotope-independent, and $M_A$ is the nuclear mass for atomic number $A$. Higher order corrections are smaller than the current experimental and theoretical uncertainties for the cases studied in this work. 
The MS constant is usually separated into $K=K_{\text{NMS}}+K_{\text{SMS}}$, with $K_{\text{NMS}}$ and $K_{\text{SMS}}$ the normal mass shift (NMS) and the specific mass shift (SMS) constants, respectively.
For atomic transitions in light systems, $K_{\text{NMS}}$ can be estimated with a few per-mil precision by scaling the experimental excitation energy ($E^{ex}$) as $K_{\text{NMS}}=E^{ex} m_e$, where $m_e$ is the electron mass \cite{Bethe1977,2003-Tup,2007-Ba,1998-NaIonizationPot,1991-MgIonization,2012-NMS,2010-LiLike}. However, accurate calculations of $K_{\text{SMS}}$ are extremely challenging for systems with more than six electrons \cite{2013-Born, 2015-ECGb, 2016-B, 2019-Boron, 2020-Ar, 2020-C, 2021-Be}. When given, the reported $K_{\text{SMS}}$ uncertainties in many-electron systems are typically larger than $10\%$ \cite{2012-CalcReview, 2014-AgSMS, 2016-Mn, 2017-SMS, Sahoo_2020, Koszorus2021, 2021-Al}, thus limiting the effectiveness of comparisons between experimental charge radii with those calculated using nuclear theory.

\textit{Atomic theory.---} 
We employ the relativistic coupled-cluster (RCC) theory, which is well suited for the accurate evaluation of correlations in many-electron atomic systems \cite{Lindgren}.
Traditionally, two procedures have been used to carry out these calculations; finite-field (FF) \cite{2003-Ber,2003-Tup,2007-Korol,2015-Roy} and expectation-value-evaluation (EVE) \cite{2001-Saf,2010-Sahoo}, using many-body methods including the RCC theory \cite{2010-Sahoo,2015-Roy}. These proved to have several limitations as detailed in Refs.~\cite{2020-InRCC1, 2021-ARCCSDT}.
The analytic response-based RCC (AR-RCC) theory was developed to circumvent these problems.

Recently, we have used the AR-RCC theory with singles and doubles approximation (AR-RCCSD) to estimate the IS constants for transitions in the indium atom \cite{2020-InRCC1,Sahoo_2020}, but calculations for Na and Mg$^+$ require further development to precisely estimate these constants by including higher-order electron correlations. Thus we extend our AR-RCC theory to account for full triples excitations (AR-RCCSDT method).
This method was recently bench-marked by performing extensive calculations on lithium-like systems, for which more accurate methods are available, and was found to be the most reliable over the calculations carried out adopting the FF and EVE approaches~\cite{2021-ARCCSDT}.
To validate the method for many-electron systems, we compare the calculated and measured values in Mg$^+$, which has a similar electronic structure to Na. For Na, we develop a hybrid method based on a comparison of the neutron skin thickness deduced from the matter-radii measurements and those calculated by applying nuclear many-body theory.

\begin{table}[t]
\caption{Comparison of IS constants, $F$ in (MHz/fm$^2)$, and $K_{\text{SMS}}$ in (GHz amu), for the D1 and D2 lines from the AR-RCCSD/T methods with literature values. The experimental $F$ in Mg$^+$ is taken from a semi-empirical method \cite{2012-Yor}. The experimental $K_{\text{SMS}}$ in Mg$^+$ is derived from the results of \cite{2009-MgTrap} by subtracting the NMS and FS contributions. 
}
\begin{ruledtabular}
\begin{tabular}{l llll}
   &  \multicolumn{2}{c}{\textbf{D1}}  &  \multicolumn{2}{c}{\textbf{D2}}  \\
   \cline{2-3} \cline{4-5}
 \hline 
 $F$  & Na & Mg$^+$ & Na & Mg$^+$\\
 ~~ AR-RCCSD  & $-38.9   $ & $-126.1   $ & $-38.9   $ & $-126.1   $ \\
~~ AR-RCCSDT  & $-39.2(3)$ & $-126.3(7)$ & $-39.2(3)$ & $-126.3(7)$ \\
 ~~ Ref. \cite{2003-Tup}    &  $-36.45$  & $-123.2$ &  &  \\
 ~~ Ref. \cite{2003-Ber}    &  $-39$ &  & $-39$  & $-127$  \\
 ~~ Ref. \cite{2007-Korol}  & $-33$  & $-127$& $-33$  & $-127$ \\
 ~~ Ref. \cite{2001-Saf}     &  $-38.42$ &$-125.81$ & $-38.43$   & $-125.82$ \\
 ~~ Ref. \cite{2010-Sahoo}  & $-38.76$ & $-126.22$& $-38.80$ & $-126.32$  \\
 ~~ Exp. \cite{2012-Yor}   &  & $-127(12)$  &  &  \\
 \hline 
 K$_{\text{SMS}}$  & Na & Mg$^+$ & Na & Mg$^+$ \\
 ~~ AR-RCCSD & $132   $ & $404$ & $132     $ & $404$ \\
~~ AR-RCCSDT & $109(3)$ & $374(7)$ & $109(3)$ & $374(7)$ \\
  ~~ $\Delta$NNLO$_\mathrm{GO}$+$r_m$  & $105.3(1.3)$ &  &  & \\
   ~~ Ref. \cite{2003-Tup}    & $~98.5$ &  $406.1$  &  &  \\
  ~~ Ref. \cite{2003-Ber}    &  $109(24)$ & $379(12)$ & $108(24)$ & $373(6)$ \\
  ~~ Ref. \cite{2007-Korol}  & $116$ & 378 & $116$ & 378  \\
  ~~ Ref. \cite{2015-Roy}    &  & $365$ &  & $366$  \\
  ~~ Ref. \cite{2001-Saf}     & ~$97$ & 362 & ~$97$ & 361    \\
  ~~ Ref. \cite{2010-Sahoo}  & $114.4$ & 398.8 & $112.3$ & 389.9 \\
 ~~ Exp. \cite{2009-MgTrap}    &  & $369.3(3)$  &  & $367.7(3)$  \\

\end{tabular}
\end{ruledtabular}
\label{natab}
\end{table}

\textit{Nuclear theory.---}
Accurate \textit{ab initio} calculations of charge radii are particularly challenging for open-shell nuclei. Shell-model calculations based on  non-perturbative effective interactions derived from methods like valence-space in-medium similarity renormalization group~\cite{hergert2016,ragner2017,miyagi2020} and shell-model based coupled-cluster~\cite{sun2018}, are complicated for nuclei where valence spaces consist of more than one major shell. Alternatively, methods based on single reference states that explicitly break symmetries may provide a conceptually simpler approach~\cite{soma2020,sam2020,Soma2021}. 
However, such approaches carry uncertainties from the lack of symmetry restoration that are somewhat difficult to quantify. In this work, we follow Ref.~\cite{sam2020} and employ single-reference coupled-cluster (CC) theory.

\begin{table*}[t]

\caption{
Ground states properties of the Na isotopic chain.
The first two columns give the mass and neutron numbers.
Columns 3 and 4 give the experimental charge radii differences, and absolute values respectively, for $K_{D1}=388(3)$ and $F_{D1}=-39.2(3)$, with uncorrelated uncertainties in parenthesis, and correlated ones from our estimation of $K$ are in square brackets.
Columns 5-7 give the charge, proton, neutron radii and skin, respectively, while the neutron skin is given in the last column, as calculated via $\Delta$NNLO(450).
}
\label{NAR}
 \begin{ruledtabular}
    \begin{tabular}{cclccccc}
$A$ & $N$ & $(\delta r_c^2)^{23,A}\,$fm$^2$ &  $r_c\,$fm & $r_c^{Th}\,$fm  & $r_p^{Th}\,$fm  & $r_n^{Th}\,$fm  & $r_{np}^{Th}\,$fm 
\\
 \hline \\
19 & ~8 &   &                                 & 2.99(3) & 2.87(3)   & 2.58(3) & -0.29(3)\\
20 & ~9 &  -0.60(9)[52] & 2.891(16)[92]     &   \\
21 & 10 &  -0.13(5)[33] &   2.972~(8)[57]     & 3.03(3) & 2.92(3)   & 2.82(3) & -0.09(3)\\
22 & 11 &   -0.16(2)[16] &   2.967~(3)[27]    &  \\
23 & 12 &                  &    2.9935(38) ~ ~& 3.01(3) & 2.90(3)   & 2.92(3) &  0.02(3)\\
24 & 13 &-0.02(4)[15] &   2.990~(7)[25]       &  \\
25 & 14 & ~0.12(5)[28] &   3.013~(8)[47]      & 2.97(3) & 2.87(3)   & 2.98(3) &  0.11(3)\\
26 & 15 &   ~0.34(2)[41] &   3.049~(4)[68]    &  \\
27 & 16 &  ~0.60(5)[52] &   3.091~(8)[86]     & 3.01(3) & 2.92(3)   & 3.11(3) & 0.19(3) \\
28 & 17 &  ~0.89(7)[63] &  ~ 3.139(11)[101]    &  \\
29 & 18 &  ~1.39(10)[73] &   ~ 3.217(15)[114]  & 3.03(3) & 2.94(3)   & 3.25(3) & 0.31(3) \\
30 & 19 &   ~1.68(15)[82] & ~ 3.262(23)[128]  &  \\
31 & 20 & ~2.19(9)[91]    & ~  3.339(13)[137] & 3.06(3) & 2.96(3)   & 3.33(3) & 0.37(3)  \\%
33 & 21 &     &                               & 3.13(3) & 3.04(3)   & 3.49(3) & 0.45(3) \\%
    \end{tabular}
    \end{ruledtabular}
    \label{tab:results}
\end{table*}


For the CC calculations we employ the recently developed $\Delta$NNLO$_{\mathrm{GO}}$ interaction~\cite{jiang2020} with a momentum cutoff of 450~MeV.
These calculations are performed in the singles and doubles (CCSD) approximation~\cite{kuemmel1978,2007-Bartlett,hagen2014} starting from an axially symmetric Hartree-Fock (HF) reference state. Parity, particle number, and the projection of total angular momentum onto the symmetry axis are conserved quantities. The HF calculations are performed in a harmonic-oscillator basis consisting of 15 major oscillator shells ($N_{\rm{max}} = 14$), with a spacing of $\hbar\omega=16$~MeV. The three-body interaction has an additional energy cut given by $E_{\rm{3max}}=16$~MeV, which is sufficiently large for the nuclei we compute. Once the HF solution is converged, a more accurate density matrix is computed using second-order many-body perturbation theory~\cite{tichai2019}. Diagonalization of this density matrix then yields the natural orbital basis~\cite{tichai2019,sam2020,hoppe2021}. Following Refs.~\cite{hoppe2021,heinz2021} the normal-ordered Hamiltonian in the two-body approximation~\cite{hagen2007a, roth2012} is then truncated to a smaller model space ($N_{\rm{max}}^{\rm{nat}} =12$) according to the occupation numbers of the natural orbits. 
The proton and neutron radii are calculated as ground-state expectation values, and charge radii include corrections from the Darwin-Foldy term and spin-orbit contributions (see Supplemental Material for details). So far, we can only estimate the effects of symmetry restoration from projected HF calculations. The uncertainties of nuclear radii from model-space truncation and uncertainties from the interaction are estimated to be 2$-$3\% following Refs.~\cite{sam2020,Koszorus2021}.  

Odd-mass nuclei, such as the Na isotopes considered in this work, are more complicated than even-even nuclei because of the unpaired last nucleon. We performed quadrupole constrained HF calculations for a range of oblate and prolate deformations and found that in all cases (except for $^{25}$Na) the prolate HF minimum provides the optimal reference state for the CC calculations. For $^{25}$Na, starting from an oblate reference state yield the largest binding energy.

{\it Results and discussion.---} Table~\ref{natab} shows  our results of $F$ and $K_{\text{SMS}}$, computed with the AR-RCCSD and AR-RCCSDT methods, for the D1 and D2 atomic transition lines of Na and Mg$^+$. The calculated energies are compared with the experimental values in the Supplemental Material. We find that by including triples excitations, the energy calculation accuracy is improved by an order of magnitude.
The differences between the AR-RCCSD and AR-RCCSDT values for $F$ in all the states are found to be negligible. This finding is in line with our calculations for Li-like systems \cite{2021-ARCCSDT}, where it was also found that both the EVE and AR methods produce reliable results for $F$, with only the FF method showing some spurious deviations. These two facts explain the agreement between our calculation and the previously reported FF \cite{2003-Ber,2015-Roy} and EVE \cite{2010-Sahoo,2001-Saf} results from the literature. The FF results of \cite{2003-Tup,2007-Korol} show a more significant deviation. Our uncertainty for $F$ is given by estimating the magnitude of neglected QED effects as detailed in Ref.~\cite{2021-ARCCSDT}.

In contrast to the results for calculations of $F$, we find triples excitations to be significant for the $K_{\text{SMS}}$ constants in both systems, with their magnitude much larger than in Li-like systems \cite{2021-ARCCSDT}. 
An indication to this behaviour comes from the fact that in both Na and Mg$^+$, the value of $K_{\text{SMS}}$ for the ground state changes sign between the results from the mean-field Dirac-Hartree-Fock (DHF) and AR-RCCSD/T methods (see Supplemental Material).
This highlights the critical role of electron correlations in the determination of the above constants. On the other hand, higher-order relativistic effects are found to be small, thus strengthening our assumption that $K_{\text{NMS}}$ may be taken from the scaling-law with sufficient accuracy.
Both the limited reliability of the FF and EVE approaches for estimating $K_{\text{SMS}}$ \cite{2021-ARCCSDT, 2021-CaFF}, combined with the major role of triples excitations, which are implemented fully here for the first time, account for the major differences found between the earlier reported values of $K_{\text{SMS}}$ in these systems. Only the values of \cite{2003-Ber}, who utilized the FF approach in a non-relativistic calculation, agrees with our calculations for both systems. However their values for the individual levels differ considerably, as shown in the Supplemental Material.

Even though higher-level, such as quadruple excitations, do not contribute directly in the AR-RCC theory owing to the one-body and two-body forms of the FS and SMS operators, their inclusion can change the amplitudes of the unperturbed wave operators, causing an indirect modification to our results. An uncertainty of $2-3\%$ was estimated by analyzing such contributions in a perturbative approach. For Na, this uncertainty is an order of magnitude smaller than that reported in the literature~\cite{2003-Ber}.
To validate the reliability of our calculated $K_\mathrm{SMS}$, and their uncertainty, we carry out two benchmark tests; one from the atomic physics side, and the other from the side of nuclear physics.

\begin{figure}[t]
 \centering
\includegraphics[width=1\columnwidth]{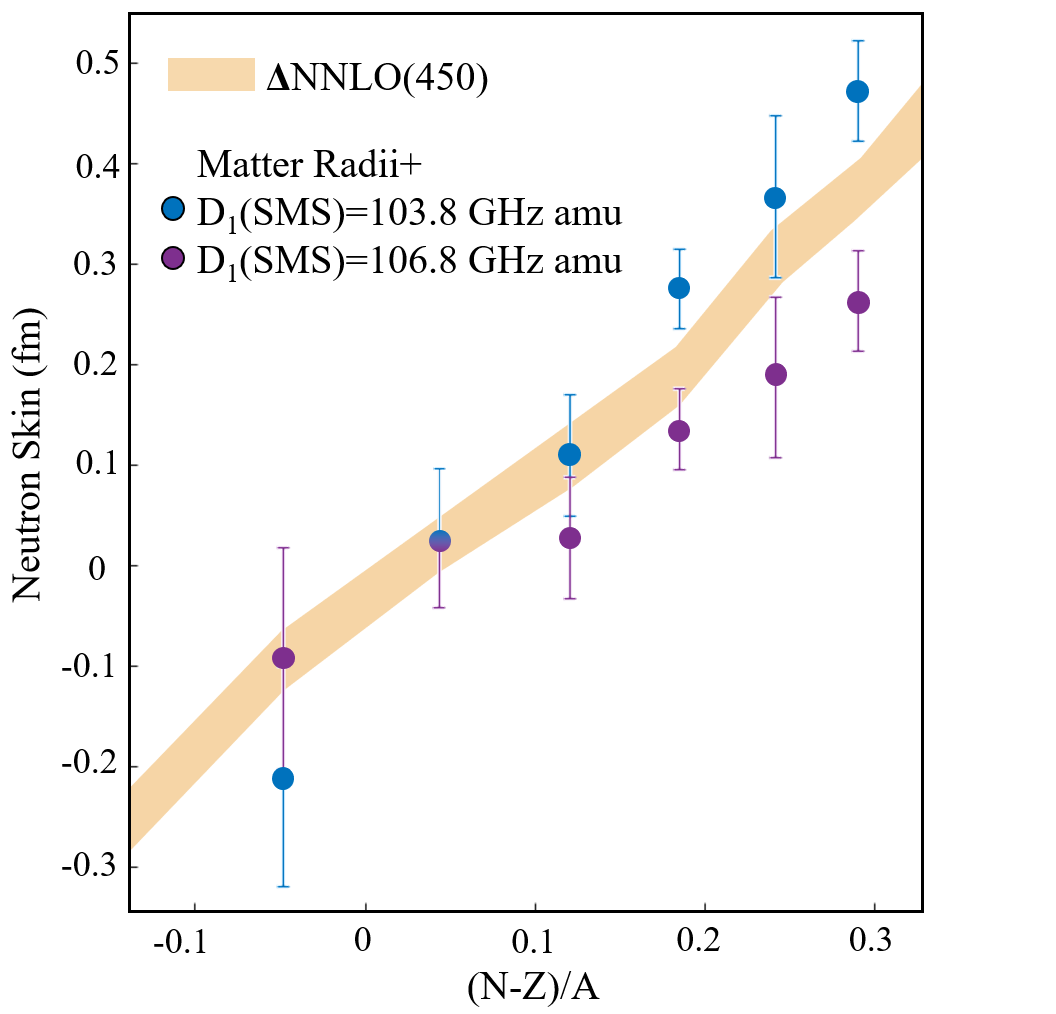}
\caption{\label{skin}
Semi-empirical estimation of $K_\mathrm{SMS}$.
We compare the calculated $r_{np}$ from Table \ref{tab:results} (filled band), with $r_{np}$ deduced from: matter radii (taken from \cite{Su98,2014-NaRM}), the field shift given in this work, and calculated spin orbit corrections.
The comparison is done for different values of $K_\mathrm{SMS}$, and returns: $K_\mathrm{SMS}=105.3(1.3)\,$GHz amu. 
} 
\end{figure}

{\it Benchmark with Mg$^+$.---}
We take advantage of the precise IS measurements for Mg$^+$ \cite{2009-MgTrap}.
Applying Eq. (\ref{eq:KP}), with $K_\mathrm{NMS}$ taken from the scaling law, and $F=-126.3(7)$ MHz/fm$^2$ from this work, as well as $(\delta r_c)^2_{24,26}=-0.158(9)$ fm$^2$ (see Supplemental Material), we obtain $K_{\text{SMS}}=369.3(3)$ GHz amu and $K_{\text{SMS}}=367.7(3)$ GHz amu for the D1 and D2 transitions, respectively. These values agree within one standard deviation with our calculated values for $K_{\text{SMS}}=374(7)$ GHz amu for both transitions.

{\it Benchmark with matter radii.---}
In Table \ref{tab:results} we give the result for the ground state properties for Na isotopes calculated by $\Delta$NNLO$_\mathrm{GO}(450)$. Our uncertainty estimation is based on similar calculations in Ne, Mg, K, and Ca for the neutron skin~\cite{hagen2015} and radii~\cite{sam2020,Koszorus2021}. 

Instead of comparing directly with the matter radii, the neutron skin, defined as the difference between its proton and neutron radii, $r_{np}=r_n-r_p$, can be used to establish physical limits for the atomic parameters.
Figure~\ref{skin} shows the nuclear theory results for $r_{np}$ alongside the semi-empirical results obtained using available matter radii~\cite{Su98,2014-NaRM}, our calculated $F=-39.2(3)$ MHz/fm$^2$ and spin-orbit corrections, and charge radii values extracted from the IS data using different values of the atomic parameter $K_\mathrm{SMS}$. 
As $r_c$ is extremely sensitive to $K_\mathrm{SMS}$, we vary this parameter, while keeping $F$ constant, and compare the extracted neutron skin with the accurate theoretical prediction.
This semi-empirical fit returns $K_\mathrm{SMS}=105.3(1.3)\,$GHz amu, which agrees with the direct calculation using AR-CCSDT of $K_\mathrm{SMS}=109(3)\,$GHz amu. A comparison with previous calculations is given in Fig. \ref{Fig:SMS}.

\begin{figure}[t]
  \centering
\includegraphics[width=\columnwidth]{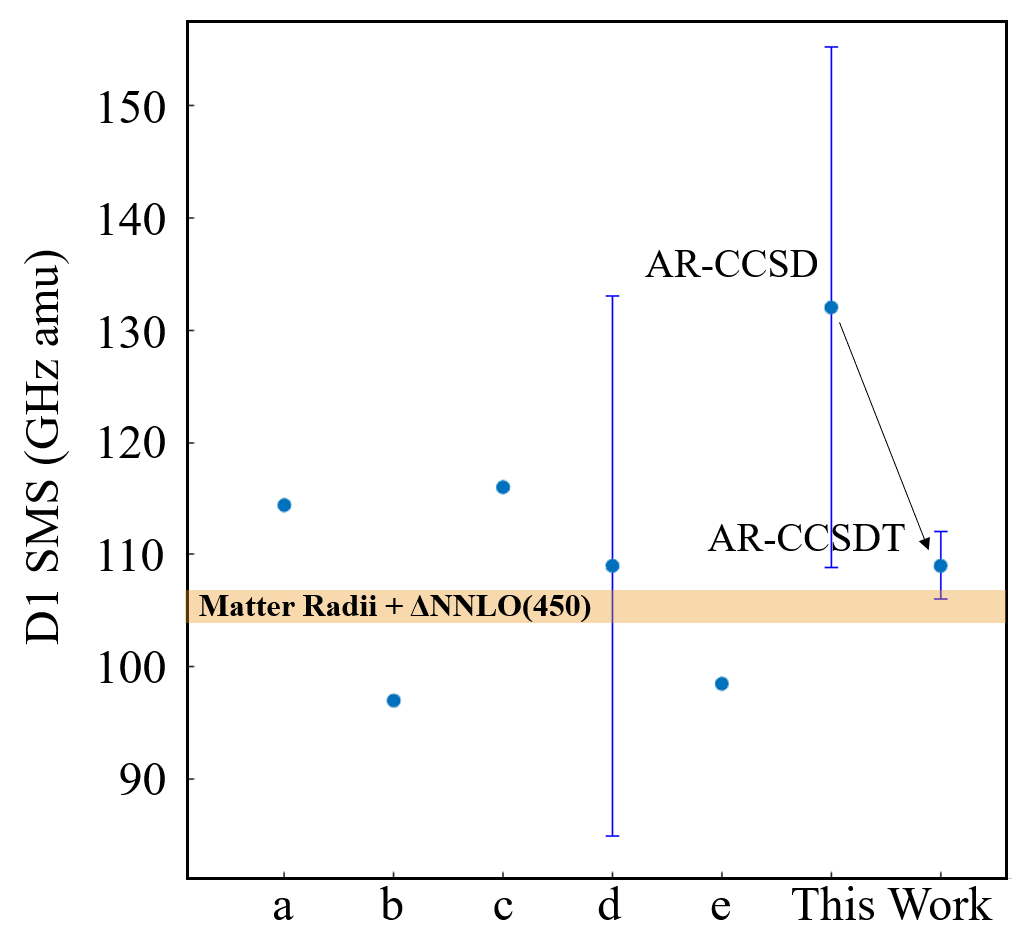}
\caption{
Comparison of $K_\mathrm{SMS}$ in the D1 line of Na as calculated by: a. \cite{2010-Sahoo}, b. \cite{2001-Saf}, c. \cite{2007-Korol}, d. \cite{2003-Ber}, e. \cite{2003-Tup}, and in this work.
The filled area is from the semi-empirical estimation illustrated in Fig. \ref{skin}.
}\label{Fig:SMS}
\end{figure}

\begin{figure*}
  \centering
\includegraphics[width=\textwidth]{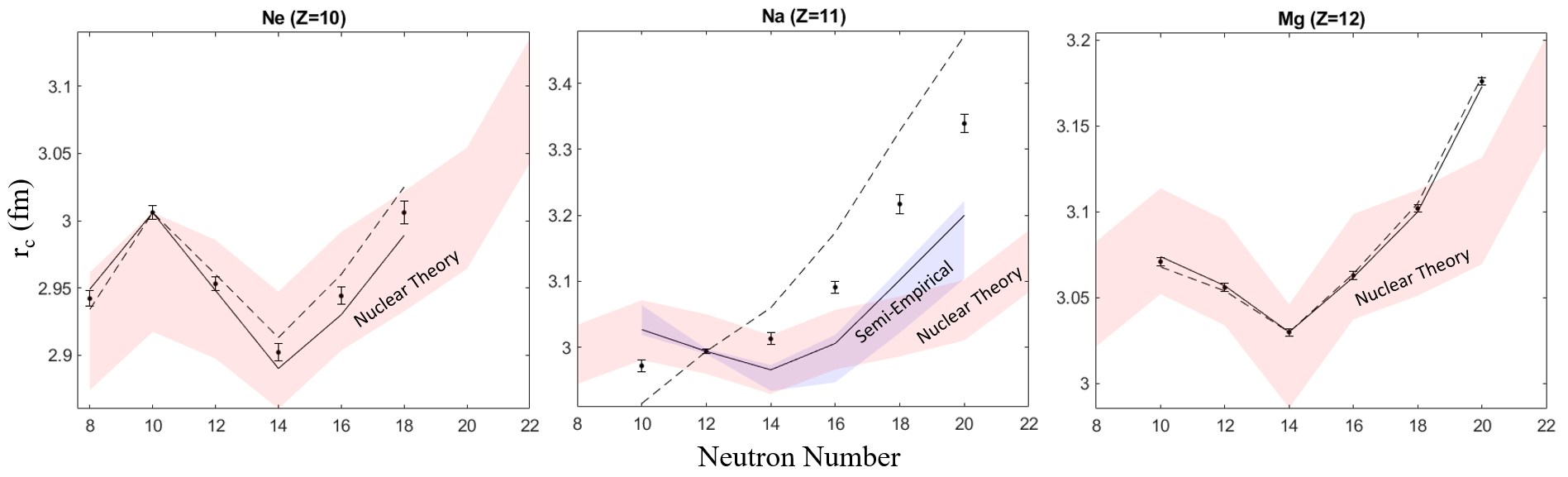}
\caption{
Absolute charge radii of Ne, Na and Mg as a function of neutron number. The data-points from \cite{2011-Ne,1975-Na1,1977-22Na,1978-Na2,1982-Na,2012-Yor} give the central values with errorbars from statistical uncertainty including that of the reference isotope. The full and broken lines give the spread resulting from the atomic parameters calculated in this work and in \cite{ohayon2019}. The filled bands represent the spread of values calculated by the nuclear theory. For Na, another band gives the values returned from the semi-empirical neutron-skin fit of Fig. \ref{skin}.
}\label{radii}
\end{figure*}

These two benchmarks validate the reliability of the central values and uncertainties calculated by the atomic theory, and serve as a striking test of the AR-CCSDT method and its applicability in light many-electron systems. 

{\it Extracted charge radii.---}
With the Na atomic constants calculated in this work (Table \ref{natab}), and the measured IS by \cite{1975-Na1,1977-22Na,1978-Na2,1982-Na}, we deduce $(\delta r_c^2)^{23,A}$ directly from Eq.~(\ref{eq:KP}).
Our results are given in Table \ref{tab:results}, and shown in Fig.~\ref{radii} along with predictions from nuclear theory.
For completeness, we also include the charge radii of Ne \cite{2011-Ne,ohayon2019} and Mg, with the latter extracted form IS
measurements \cite{2012-Yor}, and using our calculated $F=-126.3(7)\,$MHz/fm$^2$ (see Supplemental Material for more details).
Overall, nuclear theory predictions for $r_c$ agree very well with experiment for both Ne and Mg, from $N=10$ up to $N=18$. 
This agreement suggests that the $K_\mathrm{SMS}$ value expected for Na isotopes sits at the lower edge of the confidence interval given by our atomic calculations.
This conclusion is very much in line with the semi-empirical prediction from the matter radii (neutron skins), as well as the difference between experimental and calculated $K_\mathrm{SMS}$ in Mg$^+$.  Further developments in atomic theory are needed to increase the precision in the calculated atomic parameters, e.g. by including a relativistic calculation of $K_\mathrm{NMS}$, accounting for QED effects, and including quadruple excitations.

{\it Summary and outlook.---}
We report new developments of atomic theory that enable the extraction of nuclear charge radii of Na isotopes from existing isotope shift measurements, as well as a state-of-the-art \textit{ab initio} nuclear calculation of the ground-state properties of odd-mass Na isotopes.
Our results were combined with the experimental data of matter radii to extract the neutron skin values of these isotopes and establish semi-empirical constraints to the atomic physics parameters. These results provide an important benchmark to guide the development of atomic theory as currently both charge radii and neutron skin uncertainties of Na isotopes are dominated by the calculated atomic parameters.
Nuclear theory predicts a clear ``kink" in the increase of charge radii at $N=20$ and a monotonous increase of the neutron skin. As there is limited experimental data in this region, these predictions provide a further motivation to perform new experiments in these neutron-rich systems. Similar calculations are needed for other odd-$Z$ elements, for which the atomic theory input is critical to extract nuclear charge radii values from isotope shift data~\cite{2021-Al}.  These developments are timely with the progress of new radioactive beam facilities, such as FRIB in the US, where exotic light nuclear systems will be produced up to the proton and neutron drip lines.

\begin{acknowledgments}
The atomic physics calculations were carried out using the Vikram-100 HPC cluster of Physical Research Laboratory, Ahmedabad, India.
This work is supported by Department of Energy, Office of Science, Office of Nuclear Physics, under Award numbers DE-SC0021176, DE-SC0021179, DE-FG02-96ER40963 and DE-SC0018223. Computer time was provided by the Innovative and Novel Computational Impact on Theory and Experiment (INCITE) programme. This research used resources of the Oak Ridge Leadership Computing Facility located at Oak Ridge National Laboratory, which is supported by the Office of Science of the Department of Energy under contract No. DE-AC05-00OR22725.
\end{acknowledgments}

\bibliography{references,master}

\clearpage
\newpage
\maketitle

\begin{center}
\textbf{\large Nuclear charge radii of Na isotopes: A tale of two theories} \\
\vspace{0.05in}
{ \it \large Supplemental Material}\\
\end{center}

\subsection{Calculations of energies}

To validate the calculations using the employed methods, and check the accuracy of the wave functions, we present the electron affinities (opposite of ionization energies) of the $3s ~ ^2S_{1/2}$,  $3p ~ ^2P_{1/2}$ and $3p ~ ^2P_{3/2}$ states of Na and Mg$^+$ in Table \ref{engtab}. The calculations are done assuming infinite nuclear mass, and then corrected using the IS constants. We present the calculations from the DHF method, and approximating the RCC theory at the singles and doubles excitations (RCCSD method) and at the singles, doubles and triples excitations (RCCSDT method). The Breit and QED contributions from the RCCSDT method are also listed separately. The final results are compared with the experimental values \cite{1998-NaIonizationPot,1991-MgIonization}.
Compared with experiment, our calculated energies are accurate to few $10^{-4}$, with triples excitations crucial to reach this accuracy. We find higher order relativistic and recoil corrections to be negligible in this level of accuracy.

\begin{table}[h]
\caption{\label{engtab}
Calculated electron affinities (in cm$^{-1}$) of Na and Mg$^+$ using the DHF, RCCSD and RCCSDT methods. Corrections due to the Breit and QED interactions are given separately, as well as the finite mass recoil correction (NMS+SMS) determined in this work. The final results are compared with the experimental values.}
\begin{ruledtabular}
\begin{tabular}{l cc c }
 ~~ Method &  $3s ~ ^2S_{1/2}$  & $3p ~ ^2P_{1/2}$ & 	$3p ~ ^2P_{3/2}$  \\
 \hline \\
 \multicolumn{4}{l}{\bf $^{23}$Na} \\
~~ DHF     &  39951.56  &  24030.34  &  24014.12  \\
~~ RCCSD   &  41355.63  &  24464.16  &  24445.78  \\
~~ RCCSDT  &  41449.84  &  24498.53  & 24479.31   \\
~~ $+$Breit & ~ ~ $-1.86$  & ~ ~ $-1.42$   & ~ ~ $-0.32$   \\
~~ $+$QED  & ~ ~  $-3.56$  & ~ ~  ~ $0.21$      & ~ ~  ~ $0.12$  \\
~~ $+$Recoil  & ~ ~  $-1.09$  & ~ ~ $-0.53$      & ~ ~ $-0.53$  \\
\hline 
Final   &  41443(10) ~ ~ &  24496(7) ~ ~~  &  24478(7) ~ ~~ \\
Exp. \cite{1998-NaIonizationPot} & 41449.451(2)  & 24493.281(2)  & 24476.085(2)  \\
 \hline \hline \\
\multicolumn{4}{l}{\bf $^{24}$Mg$^+$} \\
~~ DHF     &  118824.0  &  84293.8  &  84203.5  \\
~~ RCCSD   &  121179.9  &  85545.6  &  85448.0  \\  
~~ RCCSDT  &  121267.6  &  85600.7  & 85502.1  \\
~~ $+$Breit & ~ ~ ~ $-8.2$    & ~ ~  $-8.7$   & ~ ~  $-3.3$  \\
~~ $+$QED   & ~ ~ ~  $-8.1$   & ~ ~ ~   0.8     & ~ ~ ~  0.6   \\
~~ $+$Recoil  & ~ ~ ~   $-3.0$  & ~ ~~ -1.6      & ~ ~~ -1.6  \\
\hline 
Final   &  121248(20) ~   &  85591(10)  ~   &  85497(10) ~   \\
Exp.\cite{1991-MgIonization}  &  121267.64(5)  &  85598.33(5)  &  85506.76(5)  \\
\end{tabular}
\end{ruledtabular}
\label{naistab}
\end{table}

\subsection{IS constant for individual states}

In the main text we give the IS constants for relevant transitions. These are taken from the differences between the constants for the individual levels, given here.

In Tables \ref{naistab} and \ref{mgistab}, we present the $F$ and K$_{\text{SMS}}$ values of the $3s ~ ^2S_{1/2}$, $3p ~ ^2P_{1/2}$ and $3p ~ ^2P_{3/2}$ states of Na and Mg$^+$, respectively, using the DHF, AR-RCCSD and AR-RCCSDT methods. The final quoted value is calculated using the Breit Hamiltonian. Its uncertainty is given by estimating numerical uncertainty, partial quadruple correlations, and unaccounted-for QED effects.

Our results for the individual states are compared with the literature where available, with a comparison for the IS parameters for the D1 and D2 lines given in the main text. As can be seen from the tables, previous calculations do not quote uncertainties and vary considerably.

\begin{table}[b]
\caption{IS constants of the $3s ~ ^2S_{1/2}$, $3p ~ ^2P_{1/2}$ and $3p ~ ^2P_{3/2}$ states in Na using the DHF, AR-RCCSD and AR-RCCSDT methods. Previously reported calculated values are also mentioned.}
\begin{ruledtabular}
\begin{tabular}{l rrr }
 Method &  $3s ~ ^2S_{1/2}$  & $3p ~ ^2P_{1/2}$ & 	$3p ~ ^2P_{3/2}$  \\
 \hline 
 \multicolumn{4}{c}{F (MHz/fm$^2$) } \\
 DHF    &  $-29.717$ & $-0.008$ & $-0.000$  \\
 AR-RCCSD  & $-37.317$  &  $1.591$ & $1.603$ \\
 AR-RCCSDT & $-37.762$  & 1.495  & 1.480  \\

 $+\Delta$Breit & 0.022 & 0.003  & 0.001  \\
 \hline 
 Final & $-37.7(3)$ & 1.50(3) & 1.48(3)  \\
  Ref. \cite{2001-Saf} & $-36.825$ & $1.597$  & $1.603$  \\
  Ref. \cite{2010-Sahoo} & $-37.035$ & $1.725$ & $1.765$ \\
 \hline  \hline
 \multicolumn{4}{c}{K$_{\text{SMS}}$ (GHz amu) } \\
  DHF & $-221.967$ &  $-115.735$ & $-115.537$ \\
 AR-RCCSD & $101.658$  &  $-29.765$ &  $-29.810$  \\
 AR-RCCSDT & 72.121   &  $-37.074$ &  $-36.960$  \\
 $+\Delta$Breit & 0.235 &  0.141 & 0.028\\
 \hline 
 Final & $72(3)$ &  $-37(1)$ &  $-37(1)$ \\
  Ref. \cite{2003-Ber} & 69 & $-40$  & $-39$ \\
  Ref. \cite{2001-Saf} & 53.94 & $-43.36$  & $-43.39$  \\
  Ref. \cite{2010-Sahoo} & 73.2 & $-41.2$ & $-39.1$ \\
\end{tabular}
\end{ruledtabular}
\label{naistab}
\end{table}

\begin{table}[t]
\caption{Same as Table \ref{naistab}, but for Mg$^+$}
\begin{ruledtabular}
\label{mgistab}
\begin{tabular}{l rrr}

 Method &  $3s ~ ^2S_{1/2}$  & $3p ~ ^2P_{1/2}$ & 	$3p ~ ^2P_{3/2}$  \\
 \hline 
 \multicolumn{4}{c}{F (MHz/fm$^2$) } \\
 DHF   &  $-104.568$  & $-0.059$  & $-0.000$  \\
 AR-RCCSD &  $-116.388$  & $9.836$    & $9.860$  \\
 AR-RCCSDT & $-116.686$  & $9.724$  &  9.746 \\
 $+\Delta$Breit & $0.096$  & $-0.006$ & $-0.009$  \\
 \hline 
 Final & $-116.6(7)$   & $9.72(8)$  & $9.74(8)$  \\
  Ref. \cite{2001-Saf} & $-116.01$ ~  & $9.800$ ~ & $9.811$ ~ \\
  Ref. \cite{2010-Sahoo} & $-116.102$ ~& $10.119$ ~  & $10.222$ ~~\\
 \hline  \hline 
 \multicolumn{4}{c}{K$_{\text{SMS}}$ (GHz amu) } \\

 DHF   & $-563.278$  & $-600.518$ &  $-598.181$ \\
 AR-RCCSD  &  125.451    & $-278.034$ & $-277.721$\\
 AR-RCCSDT & 69.965   & $-303.543$ & $-303.232$   \\

 $+\Delta$Breit & 1.051 & 0.471 &  0.026  \\
 \hline 
 Final & 71(3)  & $-303(5)$ &  $-303(5)$  \\
  Ref. \cite{2003-Ber} & 83.0 & $-296.0$  & $-290.0$ \\
  Ref. \cite{2001-Saf} & 38.0 & $-324.0$  & $-323.0$  \\
  Ref. \cite{2010-Sahoo} & $78.9$ & $-319.9$ & $-311.0$ \\
  Ref. \cite{2015-Roy}  & $-206.5$  & $-571.6$  & $-572.1$ 
\end{tabular}
\end{ruledtabular}
\end{table}

\subsection{Charge radii and their differences}

Neglecting contributions from higher moments, IS calculations and measurements pertain to the RMS charge radii differences between the isotopes. To connect these with absolute charge radii, one has in the very least use one absolute $r_c$. In principle, elastic electron scattering experiments offer a model-independent access to radii and other moments of the nuclear charge distribution, however, for the S-D shell nuclei, their results have large uncertainties and in many cases disagree with one another \cite{1992-FrickSD}. Muonic X-ray measurements are more precise, and agree better with one another. However, they give access to the model-independent Barrett-equivalent radius $R_k^\alpha$ of the charge distribution \cite{1970-Barrett}. Direct inference of $r_c$ from $R_k^\alpha$ must rely on a model-dependent assumption on the shape of the nuclear distribution. If this dependency is taken into account, it results in too large an uncertainty, an example for Mg is given in \cite{2006-Magda}.

To obtain precise and model-independent $r_c$, the gold standard is to combine electron scattering and muonic X-ray measurements \cite{2004-Frick}. This is done by determining the ratio between $R_k^\alpha$ and $r_c$, which reads
\begin{equation}\label{Eq:prop}
    \nu^2=\frac{r_c^2}{(3/5)R_k^{\alpha 2}}.
\end{equation}
The proportionality factor $\nu$ tends to $1$ for a hard sphere distribution, and increases slowly up to $Z=60$ \cite{2002-BArretAng}.
However, for the isotopes considered in this work, no such $\nu$ factors are available in the literature.

When their results are tabulated in a model-independent way, $\nu$ may be determined with high accuracy of order $10^{-4}$ from scattering measurements \cite{92-Maz}, as the main uncertainties tend to cancel in the ratio \cite{1995-Fricke}.
Such information is available for $^{26}$Mg \cite{1988-26Mg}, and $^{24}$Mg \cite{1974-24Mg, 1987-Vries}. The resulting $r_c^{\mu e}$ is given in Table \ref{tab:R_nu}, and its uncertainty is dominated by that of nuclear polarization corrections given in \cite{2004-Frick}.

\begin{table}[t]
\caption{Summary of electromagnetic moments of stable isotopes determined by muonic X-ray and electron scattering experiments. The structure of the given uncertainties is: (statistical)[Nuclear Polarization]\{proportionality factor\}.
 }\label{tab:R_nu}
\begin{ruledtabular}
    \begin{tabular}{cccc}
 & $R^\alpha_k$ (fm)&   $\nu$ &  $r_c^{\mu e}$ (fm)\\
 \hline \\
 $^{23}$Na & 3.8492(7)[30] &   1.004\{1\}~ &  ~2.9935(5)[23]\{30\}\\
 \\
$^{24}$Mg & 3.9291(5)[30] &   1.0040\{3\} &  3.0556(4)[23]\{9\} \\
$^{26}$Mg & 3.8992(8)[26] &   1.0031\{1\} &  3.0297(6)[20]\{3\} \\
\\
 & $\delta R^\alpha_k$ (fm) &    $(\delta r_c^{\mu e})^2$ (fm$^2$)\\
 \hline \\
$^{24,26}$Mg & -0.0299(9)[10] &    -0.158(4)[5]\{6\} \\

    \end{tabular}
    \end{ruledtabular}
\end{table}


To estimate $r_c^{\mu e}$ for $^{23}$Na, for which model-independent analysis of electron scattering measurements is not available, we interpolate between $\nu$ of $Z=6-16$ elements extracted from \cite{1987-Vries, 1995-Fricke}. To account for this interpolation, a conservative uncertainty in $\nu$ is taken. Our result is given in Table \ref{tab:R_nu}. 

   \begin{table}[!tb]
 \caption{Updated charge radii of Mg isotopes. Neutron deficient (ND) and neutron rich (NR) datasets are analyzed separately. For radii differences statistical uncertainties are in parenthesis and systematics related to $K'$ are in square brackets. For the absolute radii only the total uncertainty is tabulated.}\label{MgRad}
 \begin{ruledtabular}
    \begin{tabular}{cccc}
$A$ &  \cite{2012-Yor} &   $(\delta r_c^2)^{26,A}$ fm$^2$ &  $r_c$ fm \\
\hline \\

21 & ~0.173(24)[~63] & ~~0.223(25)[29]  & 3.0663(60)\\
22 & ~0.214(~5)[~51] & ~~0.251(~5)[23]  & 3.0709(37)\\
23 & ~0.053(~6)[~34] & ~~0.080(~6)[16]  & 3.0428(31)\\
24 & ~0.140(~5)[~25] & 0.158(~9) ~~   & 3.0556(25)\\
25 & -0.030(~4)[~11] & -0.022(~5)[~5]   & 3.0261(24)\\
26 & 0~~~~~ & 0~~~~~            & 3.0297(21)\\
27 & -0.008(~4)[~10] &  -0.016(~4)[~5]    & 3.0270(29)\\
28 & ~0.216(~9)[~27] & ~0.202(~9)[10]     & 3.0628(36)\\
29 & ~0.256(~6)[~36] & ~0.234(~6)[14]     & 3.0681(39)\\
30 & ~0.473(~5)[~56] & ~0.446(~5)[18]     & 3.1023(44)\\
31 & ~0.710(13)[~79] & ~0.676(13)[22]     & 3.1393(52)\\
32 & ~0.948(~6)[101] & ~0.909(~6)[26]     & 3.1761(53)
\end{tabular}
\end{ruledtabular}
\footnotesize{
$K'_{\text{ND, D1}}=951.3(4)$ GHz amu, $K'_{\text{NR, D1}}=952.1(4)$ GHz amu and $F_{\text{D1}}=-126.3(7)$ Mhz/fm$^2$.
}
\end{table}

\begin{table}[!tb]
\caption{Ground state nuclear radii of Na. All values are in fm.}\label{tab:NaGround}
 \begin{ruledtabular}
    \begin{tabular}{ccccc}
$A$  &     $r_c$    & $r_p$        & $r_m$     & $r_n$\\
   &    Main Text & Eq. (\ref{Eq:r_p})      &  \cite{Su98,2014-NaRM}     & Eq. (\ref{Eq:r_n})  \\
   \hline \\
20 &   2.90(9)    &              &  2.73(3)  &\\
21 &   2.98(6)    &   2.86(6)    &  2.75(3)  & 2.63(10) \\
22 &   2.97(3)    &              &  2.88(5)  &\\
23 &  ~ 2.993(4) &  ~ 2.89(4)  &  2.90(3)    &  2.91(7)\\
24 &   2.99(3)    &              &  2.84(4)  & \\
25 &   3.01(5)    &   2.92(5)    &  2.89(3)  &  2.86(7)\\
26 &   3.04(7)    &              &  2.93(4)  & \\
27 &   3.08(8)    &    3.00(9)   &  3.01(2)  &  3.01(7) \\
28 & ~  3.13(10)  &              &   3.11(2)&\\
29 & ~  3.21(11)  & ~ 3.13(12)   &  3.15(4) &    3.16(10)\\
30 &  ~  3.25(13) &              & 3.25(2)  &  \\
31 & ~  3.33(14)  & ~ 3.24(14)   & 3.31(2)  &   3.34(8) 
    \end{tabular}
    \end{ruledtabular}
\end{table}

In the main text, $(\delta r_c^{\mu e})^2$ for $^{24,26}$Mg, along with our calculated F, is used to determine mass shift constants from Eq. (\ref{eq:KP}). To extract $(\delta r_c^{\mu e})^2$, we first calculate the Barrett radii differences $\delta R$ assuming a correlation between the nuclear polarization corrections that reproduces their uncertainties as given in \cite{2004-Frick}, which are considered an upper limit. $\delta (r_c^{\mu e})^2$ are then determined from $\delta R_k^\alpha$ and $\nu$ with Eq. (\ref{Eq:prop}), accounting for all correlations. The uncertainty contribution stems from statistics, nuclear polarization and $\nu$ on similar footing. The total uncertainty in $\delta (r_c^{\mu e})^2$ given here is smaller then that used by \cite{2012-Yor} as they utilized a different procedure for determining them.

The ISs in the D1 transition of $^{21-32}$Mg$^+$ were measured at ISOLDE \cite{2012-Yor}. Combining their results with our $F=-126.3(7)$ MHz/fm$^2$, allows us to extract $(\delta r_c^2)^{26,A}$ more accurately than semi-empirically. The results are given in Table \ref{MgRad} and shown in Fig. \ref{radii} of the main text.

\subsection{Extraction of neutron skin thickness}

We detail the relevant ground state properties of Na needed to extract the neutron skin thickness, which is the difference between the point neutron and point proton RMS radii 
\begin{equation}\label{Eq:r_np}
r_{np}=r_n-r_p. 
\end{equation}
The point proton radius $r_p$ is given by \cite{2011-Rp}
\begin{equation}\label{Eq:r_p}
    r_p^2=r_c^2-R_p^2-\frac{N}{Z}R_n^2-\frac{3\hbar^2}{4m_p^2c^2}+\delta_{\text{SO}}^2,
\end{equation}
with $R_p=0.8414(19)$ fm the latest CODATA value for the RMS charge radius of the proton \cite{2019-CODATA}, and $R_n^2=-0.116(2)$ fm$^2$ that of the neutron \cite{2018-PDG}. The last term is the Darwin-Foldy correction \cite{1997-DF}.

Utilizing $r_p$ and $r_m$, $r_n$ may now be approximated as \cite{1997-rn}
\begin{equation}\label{Eq:r_n}
    r_m^2=\frac{Z}{A}r_p^2+\frac{N}{A}r_n^2.
\end{equation}
$r_{np}$ is now given by Eq. (\ref{Eq:r_np}). Its uncertainty is deduced taking into account the correlation between $r_n$ and $r_p$ which are both dependant on $r_c$. Our results for the ground-state moments in the Na chain are summarized in Table \ref{tab:NaGround}.

\end{document}